\documentclass[twocolumn,showpacs,aps,floatfix,prb]{revtex4}
\usepackage{graphicx}
\usepackage{dcolumn}
\usepackage{bm}
\usepackage{subfigure}
\usepackage{float}
\DeclareGraphicsExtensions{.eps,.png}
\begin{document}
\title{A theoretical analysis of inertia-like switching in magnets: applications to a synthetic antiferromagnet}

\author{Satadeep Bhattacharjee, Anders Bergman, Andrea Taroni, Johan Hellsvik, Biplab Sanyal and Olle Eriksson }
\affiliation{Department of Physics and Astronomy, Box 516, 75120, Uppsala University, Uppsala, Sweden}
\date{\today}

\begin{abstract}
The magnetization dynamics of a synthetic antiferromagnet subject to a short magnetic field pulse, has been studied by using a combination of first-principles and atomistic spin dynamics simulations. We observe switching phenomena on the time scale of tens of picoseconds, and inertia-like behavior in the magnetization dynamics. We explain the latter in terms of a {\it dynamic redistribution} of magnetic energy from the applied field pulse to other possible energy terms, such as the exchange interaction and the magnetic anisotropy, without invoking concepts such as inertia of an antiferromagnetic vector. We also demonstrate that such dynamics can also be observed in a ferromagnetic material where the incident field pulse pumps energy to the magnetic anisotropy.
\noindent 
\end{abstract}
\vspace{20mm}
\maketitle
\section{Introduction}
Studies of magnetization reversal by an external magnetic field is a central paradigm which relates to magnetic data storage.\cite{beaurepaire} The magnetic random access memory (MRAM) device\cite{mram} based on magneto-resistance (MR),
where magnetic bits can be described in terms high (1) or low (0) resistance of MR units,
is found to be one of the most promising memory devices for the near future. The magneto-resistance unit consists of 
a non-magnetic layer sandwiched between two magnetic metallic layers. The magneto-resistance could be either
giant magneto-resistance (GMR)\cite{gmr1,gmr2} or tunnel magneto-resistance (TMR)\cite{tmr1,tmr2} depending on the sandwiched layer, which could be 
either metallic or insulating. Such memory devices, with a very dense architecture, capable of performing with low power consumption
is proposed to be a {\it Universal memory} device with advantages of other memory devices such as static RAM (SRAM), dynamic
RAM (DRAM) and flash memory\cite{bib2}. An MRAM, which is capable of performing at the speed of SRAM
with dense architecture comparable to DRAM and non-volatility, (i.e. there is no loss of information when the 
power is switched off) is to some extent limited by it's writing speed. This speed depends on the magnetization reversal time, where the magnetization of a given magnetic unit is reversed.

The potential for applications of magnetization dynamics in technology, combined with the new knowledge of fundamental magnetic interactions in nano-sized materials, has caused a large 
increase in the interest in this field. Several mechanisms for shorter switching speeds have been proposed, 
e.g.toggle switching\cite{bib2}, all optical switching \cite{allopt} etc.
One recent suggestion is the so called inertia driven switching.\cite{ultra} Here a short magnetic field pulse, optically generated, of 100 fs, was found to cause a dynamic behavior of the magnetism of HoFeO$_3$, which went on for tens of picoseconds. In the paper of Ref. \onlinecite{ultra} it was pointed out that this behavior is expected only for materials with antiferromagnetic exchange interactions, either for bulk compounds with intrinsic antiferromagnetic exchange (e.g. like HoFeO$_3$) or in artificial magnetic tri- or multilayers, with antiferromagnetic interlayer exchange. 
It was argued that for materials with antiferromagnetic exchange interactions, 
the switching can be described by a second order differential equation of the so called anti-ferromagnetic unit vector, 
and hence have features associated to an inertia. It was in particular argued that the inertia accumulated during the short pulse, 
was responsible for the magnetization dynamics that lasted even after the pulse was switched off.

In this paper we address the microscopic mechanisms behind this kind of switching 
for a magnetic tri-layer, using atomistic spin-dynamics simulations.\cite{antropov,skubic}
We have chosen for our studies a Fe/Cr/Fe model system, with antiferromagnetic interlayer exchange interaction (IEC). A schematic figure of the system considered is shown in Fig.\ref{fig1}.
The exchange constants were calculated from first principles by mapping the ground state electronic structure to a Heisenberg Hamiltonian, 
using the Liechtenstein-Katnelson-Gubanov method.\cite{sasha} 
We have calculated the time-evolution of the magnetizations of a Fe/Cr/Fe trilayer
using spin dynamics simulation based on Landau-Lifshitz-Gilbert method, using the UppASD package.\cite{skubic} In our simulations we have analyzed the different terms that play a role in the switching behavior of materials with antiferromagnetic exchange, which are exposed to ultra short pulses. 
Although our simulations reproduce much of the observations reported in Ref.\onlinecite{ultra}, our interpretation of the microscopic mechanisms is different, 
and we argue that a more appropriate description of the mechanism behind this switching is described as due to a redistribution of the energy from the ultra fast magnetic pulse to the antiferromagnetic exchange interaction  and magnetic anisotropy of the material. After the pulse is switched off
the dynamics is governed by an effective field, now composed only of the exchange and anisotropy fields, 
that temporally evolves for much longer time scales than the initial pulse. The dynamically evolving magnetization 
carries the effect of the field pulse applied at the initial phase of the simulation and the effect persists through
the course of simulation.

\section{Method}
The temporal evolution of sub-lattice magnetizations are calculated using the dynamical equation of
Landau-Lifshitz-Gilbert form expressed in terms of atomic moments\cite{sd}
\begin{equation}
\frac{d{\bf m_i}(t)}{dt}=-\frac{\gamma}{1+\alpha^2}{\bf m_i}(t)\times [{\bf H^{i}_{eff}}+\frac{\alpha}{M_s}~({\bf m_i}(t)\times \bf H^{i}_{eff})]
\end{equation}
where ${\bf m_i}(t)$ is the atomic moment on the i$^{th}$ site at time t.
$\gamma$ is the gyro-magnetic ratio and $\alpha$ is the Gilbert damping factor. 
The effective field $\bf H^{i}_{eff}$ on the $i^{th}$ atom is calculated from the effective magnetic Hamiltonian given by
\begin{equation}
H_{Mag}=H_{ex}+H_{MA}+H_{a},
\end{equation}
through $${\bf H^{i}_{eff}}=-\frac{\partial H_{Mag}}{\partial{\bf m_{i}}(t)}.$$
The sub-lattice magnetization for a given Fe or Cr sublattice is simply $${\bf M}(t)=\sum_{i}{\bf m_i}(t).$$
The first term in the Hamiltonian describes the interatomic exchange interaction between between all atoms of 
the simulation cell. It can be separated into two parts; (i) pair-exchange between the atomic moments in the same layer
(intra-layer exchange coupling) and (ii) the exchange coupling between the moments residing in different layers (inter-layer
exchange coupling).
The second term of Eqn.2 represents the magnetic anisotropy. The form of magnetic anisotropy Hamiltonian
we have considered is $H_{MA}=Ksin^2\theta$, where $\theta$ is the angle
which the sublattice magnetization makes with anisotropy axis. K is the anisotropy energy density and can be written as 
the sum of two contributions $K=K_{mca}+K_{shape}$. Furthermore, $K_{mca}$ is the energy density which corresponds to magneto-crystalline
anisotropy and $K_{shape}$ corresponds to the energy density for the shape anisotropy arising from dipole-dipole interaction.
We choose K such a way that the hard axis is considered to be along the z-direction 
and the xy-plane is considered as easy plane, due to the shape anisotropy of thin magnetic films and due to the possible 
intrinsic uniaxial anisotropy.
The last term of Eqn.2 is interaction due to the applied magnetic field. For the present case it describes the 
pulsed field we apply during a short time of the simulations. Equation 1 was solved numerically using the procedure of the 
Ref. \onlinecite{mentink}.

For T= 0 K we consider the macrospin approximation, i.e. all atomistic spins within one layer are assumed to co-rotate. For finite temperature
simulations, one has to deal with this problem in an atomistic way, considering a large number of atoms in the simulations.\cite{skubic}
As mentioned, the exchange constants were calculated from first principles by mapping the ground state electronic structure to a Heisenberg Hamiltonian. 
The ground state electronic structure was calculated using the Korringa-Kohn-Rostocker Green's function method within the atomic sphere approximation.\cite{kkr1,kkr2}
In reality the interlayer exchange between the Fe layers through the
Cr spacer layer are influenced by the interface roughness of the Fe-Cr interface, and to simulate this we rescaled our interlayer exchange interactions with a factor of 10, which is according to previous analysis.\cite{erikPNAS1,erikPNAS2}

\begin{figure}[h]
\includegraphics[scale=0.4]{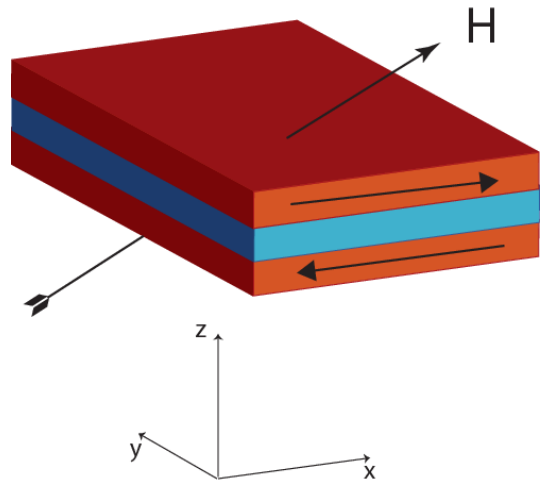}
\caption{(Color online) (Color online) The considered geometry of the atomistic simulations. 
The red slabs represent Fe and the blue slab Cr. The arrows in the slabs represent directions of magnetization.}
\label{fig1}
\end{figure}

We consider each Fe sublattice has magnetization ${\bf M}=M_s{\bf \hat{n}}(t)$, where ${\bf \hat{n}}(t)$ is the unit vector. 
The magnitude of the effective anisotropy field is given by
${F_{MA}}=\frac{2K}{M_s}cos\theta$.
$M_s$ is the saturation moment for the Fe sublattice. 
The initial value of a Fe sublattice magnetization is along the positive or negative x-direction, as shown in Fig.\ref{fig1}, i.e. ${\bf M}=\pm M_s{\bf \hat{e_x}}$.
For all simulations we used a square shaped pulse of the external field. At time t=0, we apply a field of constant value in the x-z plane
with constant value described by ${\bf{H}_a}= [H_x~ 0~ H_z] $. 
We studied the cases where the magnetization was in-plane, 
due to the shape anisotropy, without any further in-plane anisotropy. 

\section{Results}
\begin{figure}[h]
\includegraphics[scale=0.4]{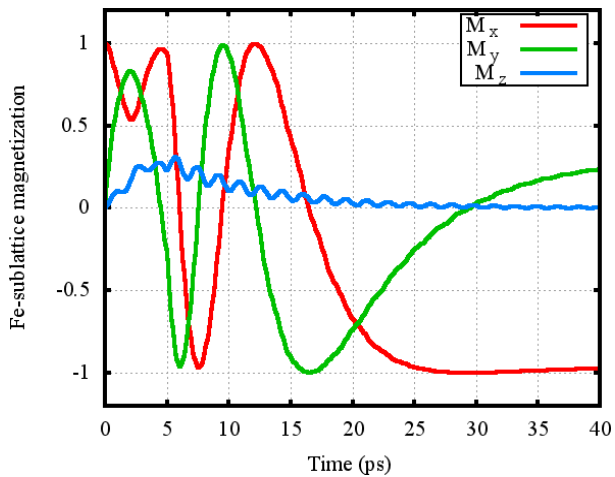}
\caption{(Color online) The time-evolution of the x-, y- and z-component of the top-most Fe sublattice 
magnetization of the tri-layer shown in Fig.\ref{fig1}. We have used H$_x$=3T, H$_z$=5T, $|$K$|$=0.06 mRy for $\alpha=0.02$}
\label{fig2}
\end{figure}

We start by showing in Fig.\ref{fig2} an example of the magnetization dynamics of the tri-layer in Fig.\ref{fig1}. Note that the magnetization shown in the figure is only for the top Fe-slab in Fig.\ref{fig1} (normalized to unity). The results shown are
for a field pulse with H$_x$= 3 T and H$_z$=5 T, for a duration of 5 picoseconds (ps). 
The Gilbert damping constant used
was 0.02 and the strength of the uniaxial anisotropy constant was 0.06 mRy/atom, providing an easy plane magnetization. 
At time t=0, we applied a field pulse.
From the figure, we see that as soon as the pulse is applied, the magnetization is brought out of plane, and all the components of magnetizations are non-zero. The magnetization then precess under the influence 
of the external field, as well as the internal fields, provided by the anisotropy field and the exchange field due to the IEC. From the figure it is clear that even after the external magnetic field is switched of (i.e. after 5 ps) the system continues to evolve in time, and the magnetization dynamics continues over the entire time-interval shown in the figure. 
At sufficiently long simulation time it seems however that the x-component of the 
magnetism approaches the negative x-direction. Hence for this choice of external 
pulse a switching of the magnetism occurs, and the direction of all magnetic moments 
illustrated in Fig.\ref{fig1} have become reversed. 
The data in Fig. \ref{fig2} are similar to those demonstrated 
experimentally in Ref. \onlinecite{ultra}, in that the magnetization 
dynamics continue long after the external pulse has been first applied and then removed from the system. 

\begin{figure}[h]
\includegraphics[scale=0.50]{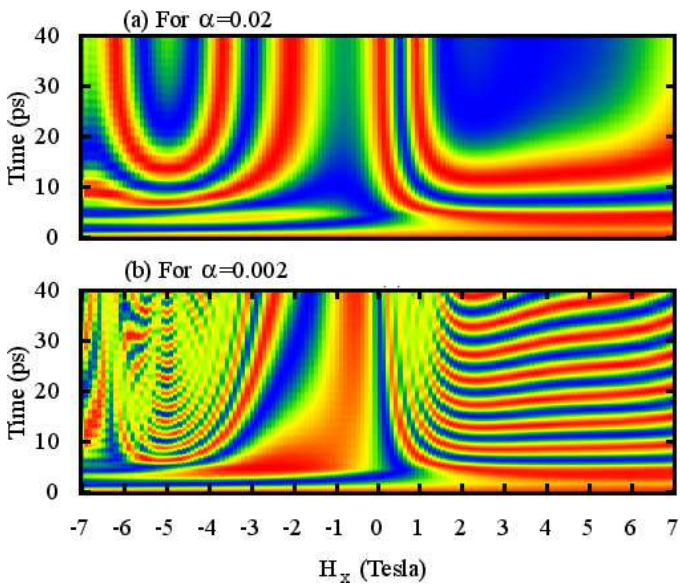}
\caption{(Color online) The switching phase diagram of an Fe/Cr/Fe-trilayer at T=0 K. The duration of the magnetic pulse is 5 ps, and the x-component of the magnetic field is shown on the x-axis, whereas the z-component is 5 T. A uniaxial anisotropy of 0.06 mRy is considered in the calculations, providing an easy plane magnetization. The x-component 
of the magnetization of the top Fe slab of Fig.\ref{fig1} is shown in a color scheme. The blue/red areas indicate the regions
where switching has occurred/not occurred (see also text for description). 
In a) we show simulations for a damping constant of 0.02 and in b) the damping constant is 0.002. H$_z$= 5T, $|$K$|$=0.06 mRy}
\label{fig3}
\end{figure}

By sweeping the strength of the x-component of the applied field we can realize different situations, in which the sublattice magnetization of one of the Fe slabs in Fig.\ref{fig1} rotates half a turn (180 degrees), a whole turn (360 degrees) or even more than one turn. In Fig.\ref{fig3} we illustrate this result, in a phase diagram of the switching behaviors. In the simulations shown in Fig.\ref{fig3} we have used the same z-component of the applied field, the same Gilbert damping factor, pulsing time, and the same uniaxial anisotropy constant, as used in Fig.\ref{fig2}. The figure only contains information about the Fe magnetization of the top Fe slab of Fig.\ref{fig1}.
In the phase diagram of Fig.\ref{fig3} the time of the simulations is shown on the y-axis and the magnitude of the x-component of the applied field is shown on the x-axis. The switched and unswitched regions are represented in terms of a color coding.
The red color specifies that the magnetization has turned $n \cdot 360$ degrees (n=0,1,2,...), whereas blue color specifies that the magnetization has turned $n \cdot 360 + 180$ degrees (n=0,1,2,...). Taking as an example H$_x$= 5T, Fig.\ref{fig3}a shows that when the pulse arrives the magnetism of the top Fe slab is along the positive x-direction (red color), but after 5-6 ps the magnetization has reversed to the negative x-direction, and after 10 ps a rotation of 360 degrees to the positive x-direction has occurred. In between, the magnetism of course points along the positive or negative y-direction (this component is not shown), and hence exhibits a rotational behavior. After some 30 ps the magnetism finally settles down along the negative x-direction. Overall, the data in Fig.\ref{fig3} shows that the magnetization dynamics continues long after the external pulse has been turned off, in a similar fashion as observed experimentally.\cite{ultra}

Another interesting feature of the data in Fig.\ref{fig3}a is that the results are not symmetric with respect to positive or negative x-component to the applied field. This might at first sight seem puzzling, since the geometry shown in Fig.\ref{fig1} is very symmetric with respect to positive and negative x-direction. However, the data in Fig.\ref{fig3}a depict only the magnetization of the topmost Fe layer, and for this layer alone there is no obvious symmetry for cases when the applied pulse is along the positive or negative x-direction, and the patterns in Fig.\ref{fig3}a for positive and negative values of H$_x$ are different. However, for all values of H$_x$ we find that after sufficiently long time the magnetism has settled down into a given orientation.

There are two torques that in general are relevant for the magnetization dynamics, a precessional torque and a damping torque, and the influence of the latter is provided in the simulations by the strength of the Gilbert damping parameter. Hence, one may expect that the 
phase diagram should depend critically on this parameter and to investigate this possibility, we show in
Fig. \ref{fig3}b a simulation with all parameters kept exactly the same as for the simulation shown in Fig.\ref{fig3}a, with the one difference that the Gilbert damping parameter was set to 0.002, i.e. ten times lower than that used for Fig.\ref{fig3}a. In this case the rotation of the magnetism continues throughout the entire period of the simulations, at least for positive values of H$_x$, which shows that it is the damping parameter that primarily determines when the magnetization stops rotating.  

\begin{figure}[h]
\includegraphics[scale=0.45]{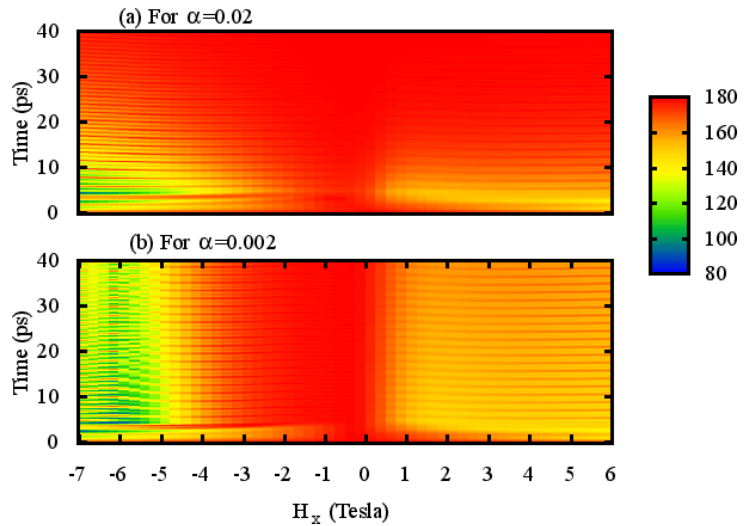}
\caption{(Color online) Evolution of angle between the sublattice magnetizations (a) for damping factor of 0.02 
(b) with a damping factor of 0.002. H$_z$= 5 T, $|$K$|$=0.06 mRy,$\alpha$=0.02}
\label{fig4}
\end{figure}

As the pulse hits the sample (simulation box) the angle between the two Fe slabs start to deviate from the 180 degree antiferromagnetic coupling. To illustrate this fact we show in
Fig.\ref{fig4} a contour plot for the angle between two Fe sublattice magnetizations, in which red coloration signifies 180 degree antiferromagnetic coupling. 
This plot corresponds to the simulations shown in Fig.\ref{fig3}a and Fig.\ref{fig3}b. 
As is clear from Fig.\ref{fig4}a,b, the magnetization of the two Fe sublattices are initially at 180 degree with each other, but after the external field is switched on
and time evolves, the angle between the sublattices 
is reduced from 180 degrees and oscillates with time, especially when H$_x\ne0$. Fig.\ref{fig4}a,b shows data for two different damping parameters, and we note that these two
different damping parameters result in slightly different magnetization dynamics, in that it takes longer time for the angle between the two sublattices to approach 180 degrees for the weakly damped system. However, for both simulations a 180 degree coupling is obtained when 
t $\rightarrow\infty$.
The minimum angle is seen when the magnitude of the x-component of the applied field is largest, after 5 ps (i.e. the duration of the pulse), where the two sublattice 
magnetizations are approximately perpendicular to each other. This means that, with the applied field as a driving force, the angle between the two sublattices is increasingly deviating, albeit in an oscillating fashion, from antiferromagnetic coupling.
When the angle between two sublattices is less than 180 degrees
an internal exchange field is build up in each Fe slab, which causes a dynamical response.  
In addition, a non-zero z-component to the magnetization (e.g. as shown in Fig.\ref{fig2}) builds up an anisotropy field. These two contributions to the effective field, which drives the magnetization dynamics, are non-zero long after the external field has been switched of. In fact, they are zero only after the magnetization has settled down to a constant value.
Hence, the effective field is in this case dynamic, and the torques due to this 
field has as usual two components; the damping part and the precessional term.

The data in Fig.\ref{fig4} demonstrate an important result, namely that during the short time 
that the external field is applied, a significant energy is established both in the exchange 
interaction term, $H_{ex}$, as well as in the anisotropy energy, $H_{MAE}$, of Eqn.2. 
In a way one can view the whole process as being due to energy provided by the $H_{a}$-term of Eqn.2, 
that during the pulse duration pumps energy into the other two terms, $H_{ex}$ and $H_{MAE}$. 
Even if $H_{a}$ is non-zero only for a short time, the other two terms will be non-zero for a much longer period, and 
the effective field composed of contributions due to these two terms depending on {\bf m$_i$}(t), carries some sort of
information of the field pulse applied at an earlier stage. 

\begin{figure}[h]
\includegraphics[scale=0.4]{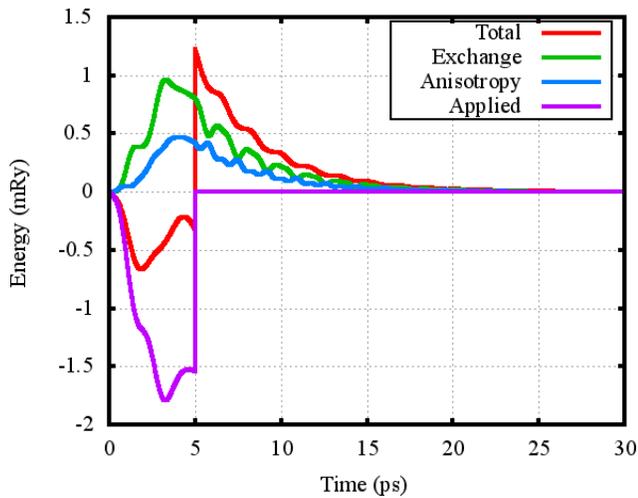}
\caption{(Color online) Time evolution of different components of magnetic energy (Exchange, Anisotropy, Applied), plotted
alongside with total magnetic energy. H$_z$= 5 T, $\alpha$=0.02 }
\label{fig5}
\end{figure}

To demonstrate this in a more formal way,
we plot in Fig.\ref{fig5} the temporal evolution of the different components 
of magnetic energy during one of the simulations. 
From Fig.\ref{fig5} one can see that the Zeeman energy is negative during the duration of the pulse whereas the
exchange and anisotropy energy are positive. The sum of the three components, i.e. the total energy is negative
during the duration of the pulse, which means that during the time that the pulse is applied, the energy is lowered 
and the system tries to relax into a new ground state configuration. When the external pulse is switched off, 
Fig.\ref{fig5} shows that there is still positive energy in the exchange and anisotropy contribution to the
energy and the total energy is now positive. In this configuration, there is both a z-component of the magnetization
as well non-collinear coupling between the moments in the two Fe-slabs. The system then relaxes back to the original 
configuration of collinear antiferromagnetic coupling of the Fe-slabs without a z-component to the magnetism and it does 
this in the anisotropy and exchange field that has been build up during the duration of the external
field . The relaxation back to collinear configuration takes considerably longer time than the duration of the field as
Fig.\ref{fig5} shows. From Fig.\ref{fig5} it can be seen that the energy 
due to applied field (Zeeman energy) acts as a reservoir for the other two terms (exchange and anisotropy) of the magnetic
Hamiltonian.
Hence, in order to analyze the magnetization dynamics provided in Figs.\ref{fig2} and Fig. \ref{fig3}, one does not need 
to invoke concepts like inertia of antiferromagnetic unit vectors\cite{ultra}. 
We suggest a simpler interpretation based on a redistribution of energy among the three terms in Eqn.2, in particular a 
redistribution of energy from $H_{a}$ to $H_{ex}$ and $H_{MAE}$, during the time period that the external field is applied. 

\par
\begin{figure}[h]
\includegraphics[scale=0.5]{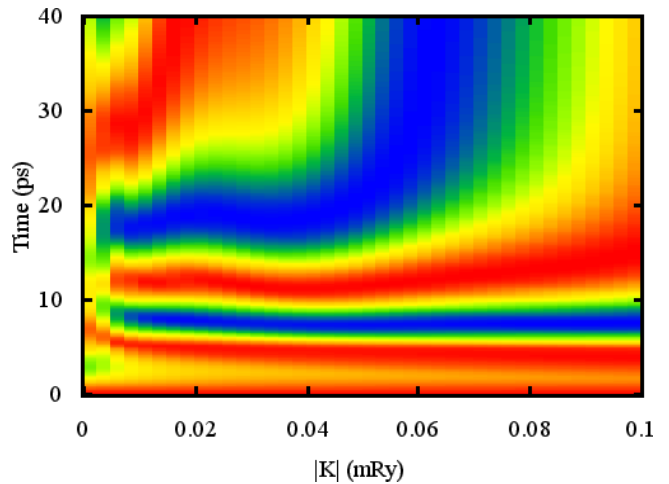}
\caption{(Color online) Switching times as function of out-of-plane anisotropy. 
The duration of the magnetic pulse is 5 ps, the z-component of the applied magnetic field is 5T and the x-component is 3T.
(All other parameters are same as Fig.2)}
\label{fig6}
\end{figure}

The analysis in the preceding paragraph indicates that also the strength of the anisotropy influenced the magnetization dynamics and to investigate this further, we plot in 
Fig.\ref{fig6}, simulated data as a function of uniaxial anisotropy constant, K. All other parameters were the same as for the simulation of Fig.\ref{fig3}a.
First we note that when K=0, the tendency of switching is reduced, since in this case the dynamics is provided by the exchange field alone, via the IEC. However, 
when K increases to 0.01-0.04 mRy, the tendency for switching increases and the magnetization is seen to precess two times before stabilizing along the positive x-direction. 
For even larger values of anisotropy energy, K larger than $\approx$ 0.05 mRy, only one revolution is found in the simulations, before the magnetization is stabilized. 
Hence there is a critical value of K which provides an optimally large amount of revolutions in the magnetization.

\begin{figure}[h]
\includegraphics[scale=0.45]{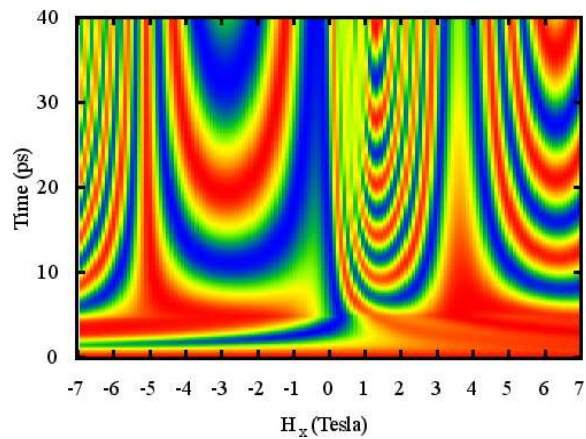}
\caption{(Color online) Switching phase diagram for Fe$_3$ layers with the same set of
parameters as indicated in Fig.3 for Fe$_3$/Cr$_4$/Fe$_3$ trilayer, except the IEC.
H$_z$= 5 T, $|$K$|$=0.06 mRy,$\alpha$=0.02}
\label{fig7}
\end{figure}

We have also investigated the case when the inter-layer exchange coupling  is zero, in Eqn.2, by performing simulations to only the top-most Fe slab, of Fig. \ref{fig1}, i.e. a purely ferromagnetic system having an easy-plane anisotropy. 
The simulated results are shown in Fig.\ref{fig7} and we note that also in this case can the magnetization behave as if it had 'inertia', i.e. it proceeds after an external field is switched off. 
An analysis relying on antiferromagnetic unit vectors becomes cumbersome in this case, but the pumping of energy from $H_a$ to the two other components given in Eqn.2, 
in this case the anisotropy energy, provides an easy interpretation. The data in Fig.7, are actually consistent with the experiments performed in Ref. \onlinecite{stamm}, where a short pulse was applied to a ferromagnet thin film.

\section{Conclusions}
In conclusion we have performed atomistic spin dynamics simulations for a synthetic antiferromagnet subjected 
to a short magnetic pulse. We have demonstrated that the magnetization dynamics continues 
long after the pulse has been applied and removed, in a fashion that is consistent with 
the experimental results of Ref. \onlinecite{ultra}. 

Several materials parameters, like the damping and strength 
of the magnetic anisotropy, are found to influence the details of the magnetization dynamics.
We suggest that such magnetization dynamics could be understood in terms of 
a redistribution of energy, during the time that the pulse is applied, 
so that the magnetic anisotropy energy and interlayer exchange energy become in an out-of-equilibrium state. 
The magnetic energy due to the incident pulse hence becomes distributed dynamically to the effective
interlayer exchange coupling between the Fe layers and it also modifies the magnetic anisotropy energy. 
The latter term ensures that the observed behavior of a magnetization dynamics that progresses long after 
an external field has been applied and switched off, is possible also in ferromagnetic materials.
\section{Acknowledgment}
We are grateful to the Swedish Research Council for support. O.E. acknowledges also support from the KAW foundation and the ERC (grant 247062 - ASD). Simulations performed on the Swedish Supercomputer Central NSC, via a grant from SNAC. Valuable discussions with the late Prof. H.-C. Siegmann, Prof. J. St\"{o}hr and A.Kirilyuk are acknowledged.

%
\end{document}